\documentclass[a4paper,11pt]{article}
\pdfoutput=1
\usepackage{jcappub}

\usepackage[utf8]{inputenc}

\newcommand{\rR}{\rho_R}
\newcommand{\rbh}{\rho_\text{BH}}
\newcommand{\nbh}{n_\text{BH}}
\newcommand{\gs}{g_\star}
\newcommand{\gss}{g_{\star s}}
\newcommand{\Tbh}{T_\text{BH}}
\newcommand{\Tbhin}{T_\text{BH}^\text{in}}
\newcommand{\Mbh}{M_\text{BH}}
\newcommand{\tev}{t_\text{ev}}
\newcommand{\Tev}{T_\text{ev}}
\newcommand{\Tbev}{\bar T_\text{ev}}
\newcommand{\Min}{M_\text{in}}
\newcommand{\nin}{n_\text{in}}
\newcommand{\Tin}{T_\text{in}}
\newcommand{\tin}{t_\text{in}}

\newcommand{\mdm}{m_\text{DM}}
\newcommand{\ndm}{n_\text{DM}}
\newcommand{\Ndm}{N_\text{DM}}
\newcommand{\gdm}{g_\text{DM}}
\newcommand{\Tp}{{T^\prime}}
\newcommand{\Tpev}{{T^\prime_\text{ev}}}
\newcommand{\Cr}{\mathcal{C}_\rho}
\newcommand{\Cn}{\mathcal{C}_n}

\title{Self-interacting Dark Matter from Primordial Black Holes}

\author[a]{Nicolás Bernal}
\author[b,\,c]{and Óscar Zapata}
\affiliation[a]{Centro de Investigaciones, Universidad Antonio Nariño\\
Carrera 3 Este \# 47A-15, Bogotá, Colombia}
\affiliation[b]{Instituto de Física, Universidad de Antioquia\\
Calle 70 \# 52-21, Apartado Aéreo 1226, Medellín, Colombia.}
\affiliation[c]{Abdus Salam International Centre for Theoretical Physics\\ Strada Costiera 11, 34151, Trieste, Italy.}

\emailAdd{nicolas.bernal@uan.edu.co}
\emailAdd{oalberto.zapata@udea.edu.co}

\abstract{
The evaporation of primordial black holes (PBH) with masses ranging from $\sim 10^{-1}$ to $\sim 10^9$~g could have generated the whole observed dark matter (DM) relic density.
It is typically assumed that after being produced, its abundance freezes and remains constant.
However, thermalization and number-changing processes in the dark sector can have a strong impact, in particular enhancing the DM population by several orders of magnitude.
Here we estimate the boost from general arguments such as the conservation of energy and entropy, independently from the underlying particle physics details of the dark sector.
Two main consequences can be highlighted:
$i)$ As the DM abundance is increased, a smaller initial energy density of PBHs is required.
$ii)$ Thermalization in the dark sector decreases the mean DM kinetic energy, relaxing the bound from structure formation and hence, allowing light DM with mass in the keV ballpark.
}

\begin{document}
\begin{flushright}
    PI/UAN-2020-680FT
\end{flushright}

\maketitle

\section{Introduction} 
The existence of dark matter (DM) has been firmly established by astrophysical and cosmological observations, although its fundamental nature remains elusive~\cite{Aghanim:2018eyx}.
Up to now, the only evidence about the existence of such a dark component is via its gravitational interactions.
In the last decades, weakly interacting massive particles (WIMPs), with masses and couplings at the electroweak scale, have been the leading DM production paradigm~\cite{Arcadi:2017kky}.
However, the increasingly strong observational constraints on DM are urging the quest for alternative scenarios.

Several alternatives to the classical WIMP mechanism exist.
For instance, one can have deviations from the standard expansion history of the early universe~\cite{Allahverdi:2020bys}.
Another possibility occurs if the couplings between the dark and visible sectors are very suppressed, so that DM never reaches chemical equilibrium with the standard model (SM), as in the case of the so-called freeze-in mechanism (FIMP)~\cite{McDonald:2001vt, Choi:2005vq, Hall:2009bx, Elahi:2014fsa, Bernal:2017kxu}.
An extreme case occurs if DM is only coupled to the SM through Planck suppressed higher dimensional operators, and is produced via purely gravitational interactions~\cite{Garny:2015sjg, Tang:2017hvq, Garny:2017kha, Bernal:2018qlk, Bernal:2020fvw, Bernal:2020yqg}. 

However, the DM genesis could also be intimately related to the Hawking evaporation of primordial black holes (PBH).
In fact, PBHs could have been formed from inhomogeneities in the early universe~\cite{Carr:1974nx}.
If their initial mass was below $\sim 10^9$~g, they disappear through Hawking evaporation~\cite{Hawking:1974sw} before Big Bang nucleosynthesis (BBN), and are poorly constrained~\cite{Carr:2009jm, Carr:2020gox}. 
During the evaporation, PBHs radiate not only SM particles but also hidden sector states, and in particular DM.
In this regard, PBH evaporation may have played a central role in the DM production~\cite{Green:1999yh, Khlopov:2004tn, Dai:2009hx, Fujita:2014hha, Allahverdi:2017sks, Lennon:2017tqq, Morrison:2018xla, Hooper:2019gtx, Chaudhuri:2020wjo, Masina:2020xhk, Baldes:2020nuv, Gondolo:2020uqv, Bernal:2020ili, Bernal:2020bjf}.

Previous studies typically assumed that after production via PBH evaporation, the DM abundance remains constant.%
\footnote{See however Ref.~\cite{Gondolo:2020uqv}, where scenarios with a second DM production mechanism have been considered.}
Nevertheless, even if there is no DM production out of the visible sector, the dynamics in the dark sector could be not trivial, featuring, for example, $N$-to-$N'$ number-changing interactions, where $N$ DM particles annihilate into $N'$ of them (with $N>N'\geq 2$).
The dominant $N$-to-$N'$ processes are naturally 3-to-2 (see, e.g., Refs.~\cite{Carlson:1992fn, Hochberg:2014dra, Bernal:2015bla, Bernal:2015lbl, Bernal:2015ova, Pappadopulo:2016pkp, Farina:2016llk, Choi:2017mkk, Chu:2017msm}), but are forbidden in models where DM is protected by a $\mathbb{Z}_2$ symmetry.
In that case, unavoidable 4-to-2 annihilations~\cite{Bernal:2015xba, Heikinheimo:2016yds, Bernal:2017mqb, Heikinheimo:2017ofk, Bernal:2018ins, Bernal:2018hjm} could dominate.
If these processes reach equilibrium, DM forms a thermal bath with a temperature in general different from the one of the SM.
More importantly, number-changing processes have a strong impact on DM, increasing by several orders of magnitude its relic abundance~\cite{Chu:2013jja, Bernal:2015ova, Bernal:2015xba, Bernal:2017mqb, Falkowski:2017uya, Herms:2018ajr, Heeba:2018wtf, Mondino:2020lsc, Bernal:2020gzm, March-Russell:2020nun}. 

In this work, we investigate the impact of thermalization of the dark sector on the DM abundance produced by evaporation of PBHs.
For that purpose, in section~\ref{sec:PBH} we briefly revisit the formation and evaporation of PBH, whereas in section~\ref{sec:DM} the standard DM production via Hawking radiation is presented.
Section~\ref{sec:SI} is devoted to quantify the DM self-interaction effects on its abundance and contains our main results.
Finally, in section~\ref{sec:con} our conclusions are presented.

\section{The Rise and Fall of PBHs} \label{sec:PBH}
Formation and evaporation of PBHs has been vastly discussed in the literature, see, for instance, Refs.~\cite{Carr:2009jm, Carr:2020gox, Masina:2020xhk, Gondolo:2020uqv}.
Here we briefly review the main aspects.

\subsection{Formation}
PBH formed in a radiation dominated epoch, when the SM plasma has a temperature $T=\Tin$, have an initial mass $\Min$ similar to the enclosed mass in the particle horizon, given by
\begin{equation}
    \Min\equiv\Mbh(\Tin)=\frac{4\pi}{3}\,\gamma\,\frac{\rR(\Tin)}{H^3(\Tin)}\,.
\end{equation}
In this expression, $\gamma\simeq w^{3/2}\simeq 0.2$ (the equation of state parameter $w=1/3$ in a radiation-dominated epoch), $\rR(T)\equiv\frac{\pi^2}{30}\,\gs(T)\,T^4$ is the SM radiation energy density with $\gs(T)$ the number of relativistic degrees of freedom contributing to $\rR$~\cite{Drees:2015exa}, and $H^2(T)=\frac{\rho(T)}{3M_P^2}$ is the squared Hubble expansion rate in terms of the total energy density $\rho(T)$, with $M_P$ the {\it reduced} Planck mass.

Extended PBH mass functions arise naturally if the PBHs are created from inflationary fluctuations or cosmological phase transitions, see, e.g., Refs.~\cite{Dolgov:1992pu, Yokoyama:1998xd, Niemeyer:1999ak, Carr:2009jm, Musco:2012au, Carr:2017jsz, Deng:2017uwc, Liu:2019lul, Carr:2020gox}.
However, for the sake of simplicity, in the present analysis, we assume that all PBHs have the same mass (i.e., they are produced at the same temperature), which is a usual assumption in the literature.
Finally, PBHs can gain mass via mergers~\cite{Zagorac:2019ekv, Hooper:2019gtx, Hooper:2020evu} and accretion~\cite{Bondi:1952ni, Nayak:2009wk, Masina:2020xhk}.
These processes are typically not very efficient, inducing a mass gain of order $\mathcal{O}(1)$, and will be hereafter ignored.

\subsection{Evaporation}
PBHs evaporate by emitting particles lighter than its temperature $\Tbh$ via Hawking radiation~\cite{Hawking:1974sw}.
Given the fact that Hawking radiation can be described as blackbody radiation (up to greybody factors), the energy spectrum of a species $j$ with $g_j$ internal degrees of freedom radiated by a nonrotating BH with zero charge is therefore~\cite{Page:1976df, Gondolo:2020uqv}
\begin{equation}
    \frac{d^2u_j(E,t)}{dt\,dE}=\frac{g_j}{8\pi^2}\frac{E^3}{e^{E/\Tbh}\pm 1}\,,\qquad \text{($+$ for fermions, $-$ for bosons)}
\end{equation}
where $u_j$ is the total radiated energy per unit area, $t$ the time, $E$ the energy of the emitted particle $j$, and
\begin{equation}
    \Tbh=\frac{M_P^2}{\Mbh}\simeq10^{13}\,{\rm GeV}\left(\frac{1\,{\rm g}}{\Mbh}\right)
\end{equation}
the BH horizon temperature.

The evolution of the BH mass due to Hawking evaporation is given by
\begin{equation}\label{eq:dMdt}
    \frac{d\Mbh}{dt} = -4\pi\,r_S^2\sum_j\int_0^\infty\frac{d^2u_j(E,t)}{dt\,dE}dE
    = -\frac{\pi\gs(\Tbh)}{480}\frac{M_P^4}{\Mbh^2}\,,
\end{equation}
where $r_S\equiv\frac{\Mbh}{4\pi\,M_P^2}$ is the Schwarzschild radius of the BH.
Assuming that $\gs$ has no temperature dependence during the whole lifetime of the BH, Eq.~\eqref{eq:dMdt} admits the analytical solution
\begin{equation}
    \Mbh(t)=\Min\left(1-\frac{t-\tin}{\tau}\right)^{1/3},
\end{equation}
where $\tin$ corresponds to the time at formation of the PBH and
\begin{equation}
    \tau\equiv\frac{160}{\pi\,\gs(\Tbh)}\frac{\Min^3}{M_P^4}
\end{equation}
is the PBH lifetime.

Complete BH evaporation happens at $t=\tev\equiv \tin+\tau\simeq\tau$ at which $\Mbh(\tev)=0$.%
\footnote{Where it has been taken into account that $\frac{\tin}{\tau} \simeq 7.86\times10^{-12} \left(\frac{\gs(\Tbh)}{106.75}\right) \left(\frac{0.2}{\gamma}\right)\left(\frac{1\,{\rm g}}{\Min}\right)^2\ll 1$.} 
In a universe dominated by radiation during the whole BH lifetime, it corresponds to a temperature
\begin{equation}\label{eq:Tev}
    \Tev \equiv T(\tev)\simeq \left(\frac{9\,\gs(\Tbh)}{10240}\right)^\frac14 \left(\frac{M_P^5}{\Min^3}\right)^\frac12 \simeq 1.2\times10^{10}~{\rm GeV} \left(\frac{\gs(\Tbh)}{106.75}\right)^\frac14\left(\frac{1~{\rm g}}{\Min}\right)^\frac32,
\end{equation}
using the fact that $H(t)=1/(2t)$ in a radiation-dominated era.

Additionally, the total number $N_j$ of the species $j$ of mass $m_j$ emitted during the PBH evaporation is
\begin{equation}\label{eq:N0}
    N_j=\int_{t(m_j)}^{\tev}dt\int_0^\infty dE\,\frac{d^2N_j}{dt\,dE}
    =\int_{t(m_j)}^{\tev}dt\int_0^\infty dE\,\frac{4\pi\,r_S^2}{E}\frac{d^2u_j}{dt\,dE}\,,
\end{equation}
where $t(m_j)$ corresponds to the time at which BH start emitting $j$ particles, i.e. when  $\Tbh\geq m_j$, and is given by
\begin{equation}
    t(m_j)=\max\left[\tin\,,\,\,\tin+\tau\left(1-\left[\frac{M_P^2}{m_j\,\Min}\right]^3\right)\right].
\end{equation}
Therefore, Eq.~\eqref{eq:N0} reduces to
\begin{equation}\label{eq:N}
    N_j=\frac{15\,\zeta(3)}{\pi^4}\frac{g_j\,\Cn}{\gs(\Tbh)}
    \begin{cases}
        \left(\frac{\Min}{M_P}\right)^2\qquad\text{for}\quad m_j\leq\Tbhin\,,\\[8pt]
        \left(\frac{M_P}{m_j}\right)^2 \qquad\text{for}\quad m_j\geq\Tbhin\,,
    \end{cases}
\end{equation}
where $\Tbhin\equiv\Tbh(t=\tin)$ is the initial PBH temperature, and $\Cn = 1$ or $3/4$ for bosonic or fermionic species, respectively.
Additionally, their mean energy is~\cite{Baumann:2007yr,Fujita:2014hha, Morrison:2018xla}
\begin{equation}\label{eq:meanE}
    \langle E_j\rangle = \frac{1}{N_j}\int_{t(m_j)}^{\tev}dt\int_0^\infty dE\,3\Tbh\,\frac{d^2N_j}{dt\,dE}=
    \begin{cases}
    6\,\Tbhin \qquad&\text{for}\quad m_j\leq\Tbhin\,,\\[8pt]
    6\,m_j \qquad&\text{for}\quad m_j\geq\Tbhin\,,
    \end{cases}
\end{equation}
where $3\Tbh$ is the average energy of particles radiated by a PBH with temperature $\Tbh$.%
\footnote{The emission is not exactly blackbody but depends upon the spin and charge of the emitted particle~\cite{Page:1976df}.}\\

PBH evaporation produces all particles, and in particular extra radiation that can modify successful BBN predictions.
To avoid it, we require PBHs to fully evaporate before BBN time, i.e., $\Tev>T_\text{BBN}\simeq 4$~MeV~\cite{Sarkar:1995dd, Kawasaki:2000en, Hannestad:2004px, DeBernardis:2008zz, deSalas:2015glj}, which translates into an upper bound on the initial PBH mass%
\footnote{The corresponding bound for BH domination decreases by a factor $(3/4)^{1/3}$.}
\begin{equation}
    \Min \lesssim 2\times 10^8~\text{g}\,.
\end{equation}
On the opposite side, a lower bound on $\Min$ can be set once the upper bound on the inflationary scale is taken into account.
The limit reported by the Planck collaboration $H_{I} \leq 2.5\times10^{-5}M_P$~\cite{Akrami:2018odb}  implies that   
\begin{equation}
    \Min\gtrsim 4\pi\,\gamma\,\frac{M_P^2}{H_I}
    \simeq 0.1~\text{g}\,.
\end{equation}\\

Before concluding this section, let us note that as BHs scale like non-relativistic matter ($\rbh \propto a^{-3}$, with $a$ being the scale factor), its energy density $\rbh$ naturally tends to dominate over the SM energy density that scales like $\rR \propto a^{-4}$.
The initial PBH energy density is usually normalized to the SM energy density at the time of formation $T=\Tin$ via the dimensionless parameter
\begin{equation}
    \beta\equiv\frac{\rbh(\Tin)}{\rR(\Tin)}=\frac{\Min\,\nin}{\rR(\Tin)}\,,
\end{equation}
where $\nin$ is the initial BH number density.
A matter-dominated era (i.e., a PBH domination) can be avoided if $\rbh\ll\rR$ at all times, or equivalently if
\begin{equation}\label{eq:betac}
    \beta\ll\frac{\Tev}{\Tin}\,.
\end{equation}

It has been recently pointed out that the production of gravitational waves induced by large-scale density perturbations underlain by PBHs could lead to a backreaction problem.
However, it could be avoided if the energy contained in gravitational waves never overtakes the one of the background universe~\cite{Papanikolaou:2020qtd}:
\begin{equation}\label{eq:GW}
    \beta < 10^{-4}\left(\frac{10^9~\text{g}}{\Min}\right)^{1/4}.
\end{equation}
Additionally, Ref.~\cite{Domenech:2020ssp} found a stronger constraint on the amount of the gravitational waves coming from BBN
\begin{equation}
    \beta < 3.3\times 10^{-8} \left(\frac{0.2}{\gamma}\right)^{1/2} \left(\frac{\gs(\Tev)}{106.75}\right)^{1/16} \left(\frac{10^4~\text{g}}{\Min}\right)^{7/8}.
\end{equation}

\section{Dark Matter Production} \label{sec:DM}

The whole observed DM relic abundance could have been Hawking radiated by PBHs.%
\footnote{For the sake of completeness, notice that even if the $s$-channel exchange of a graviton gives an irreducible contribution to the total DM relic abundance~\cite{Garny:2015sjg, Tang:2017hvq, Garny:2017kha, Bernal:2018qlk}, it will be hereafter disregarded.}
The DM production can be analytically computed in two limiting regimes where PBHs dominated or not the energy density of the universe, and will be presented in the following.

\subsection{Radiation Dominated Universe}
The DM yield $Y_\text{DM}$ is defined as the ratio of the DM number density $\ndm$ and the SM entropy density $s(T)\equiv\frac{2\pi^2}{45}\gss(T)\,T^3$, where $\gss(T)$ is the number of relativistic degrees of freedom contributing to the SM entropy~\cite{Drees:2015exa}.

In a radiation dominated universe, the DM yield produced by Hawking evaporation of PBHs can be estimated by
\begin{equation}\label{eq:YdmRD}
    Y_\text{DM} \equiv \frac{\ndm(T_0)}{s(T_0)}
    = N_\text{DM}\,\frac{\nin}{s(\Tin)}
    = \frac34\,\Ndm\,\beta\,\frac{\gs(\Tbh)}{\gss(\Tbh)}\,\frac{\Tin}{\Min}\,,
\end{equation}
with $T_0$ the SM temperature at present, and where the conservation of SM entropy was used.
Additionally, $\Ndm$ is the total number of DM particles emitted by a PBH and is given by Eq.~\eqref{eq:N}.

\subsection{Matter Dominated Universe}
Alternatively, PBHs can dominate the universe energy density before their decay.
In that case, the DM yield is instead
\begin{equation}\label{eq:YMD0}
    Y_\text{DM}\equiv\frac{\ndm(T_0)}{s(T_0)}
    =\frac{\ndm(\Tbev)}{s(\Tbev)}
    =\Ndm\frac{\nbh(\tev)}{s(\Tbev)}\,,
\end{equation}
using again the conservation of the SM entropy {\it after} the PBHs have completely evaporated, and where $\Tbev$ is the SM temperature just after the complete BH evaporation.%
\footnote{In the approximation of an {\it instantaneous} evaporation of the PBHs, the SM entropy density is violated at $t=\tev$, and therefore there is a sudden increase of the SM temperature, from $\Tev$ to $\Tbev$.}
Additionally, assuming an {\it instantaneous} evaporation of the BHs at $t=\tev\simeq\tau$, one has that
\begin{equation}
    \nbh(\tev)=\frac{\rbh(\tev)}{\Min}=\frac{3M_P^2\,H^2(\tev)}{\Min}\simeq\frac{4M_P^2}{3\Min\,\tev^2}\simeq\frac{\pi^2\gs^2(\Tbh)}{19200}\frac{M_P^{10}}{\Min^7}\,,
\end{equation}
and that
\begin{equation}\label{eq:Tb}
    \Tbev^4\simeq\frac{\gs(\Tbh)}{640} \frac{M_P^{10}}{\Min^6}\,,
\end{equation}
where the fact that in a matter-dominated universe $H(t)=2/(3t)$ was used.
Therefore, the DM yield in Eq.~\eqref{eq:YMD0} can be expressed as
\begin{equation}\label{eq:YdmMD}
    Y_\text{DM}
    \simeq \frac34\, \Ndm\, \frac{\gs(\Tbh)}{\gss(\Tbh)}\, \frac{\Tbev}{\Min}\,.
\end{equation}\\
We notice that, as expected, the DM yields in the radiation dominated (Eq.~\eqref{eq:YdmRD}) and matter dominated (Eq.~\eqref{eq:YdmMD}) eras become identical in the limit $\beta\to\Tev/\Tin$, cf. Eq.~\eqref{eq:betac}, with $\Tbev=\Tev$.

\begin{figure}
	\centering
	\includegraphics[scale=0.8]{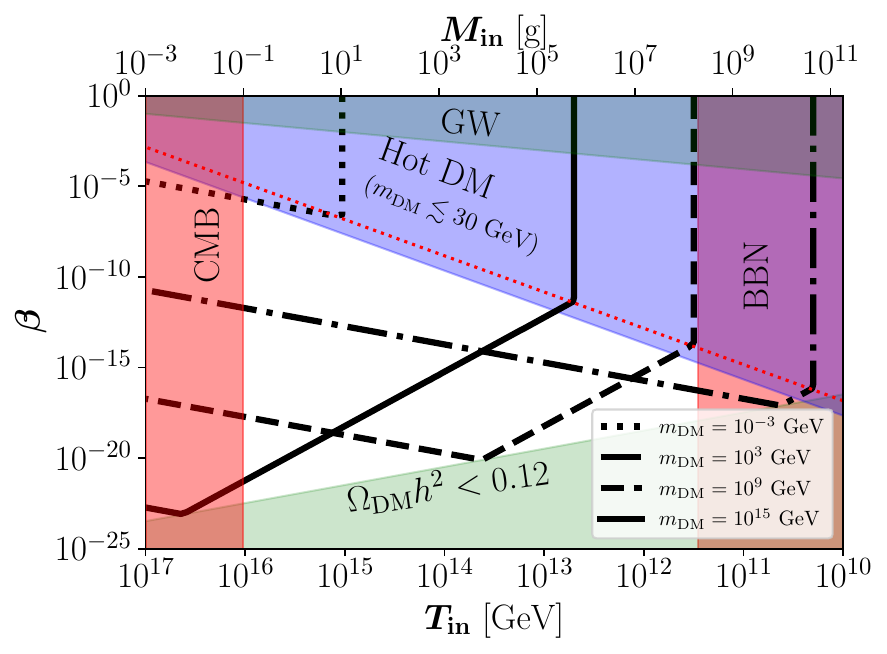}
	\caption{Parameter space reproducing the observed DM abundance (thick black lines) from PBH evaporation, {\it without} DM self-interactions.
	The shaded areas are excluded by different observables described in the text.
	The hot DM bound {\it only} applies to $\mdm \lesssim 30$~GeV.}
	\label{fig:Ti-beta}
\end{figure} 
To reproduce the observed DM relic abundance $\Omega_\text{DM} h^2\simeq 0.12$~\cite{Aghanim:2018eyx}, the DM yield has to be fixed so that $\mdm\,Y_\text{DM} = \Omega_\text{DM} h^2 \, \frac{1}{s_0}\,\frac{\rho_c}{h^2} \simeq 4.3 \times 10^{-10}$~GeV, where $\rho_c \simeq 1.1 \times 10^{-5} \, h^2$~GeV/cm$^3$ is the critical energy density, and $s_0\simeq 2.9\times 10^3$~cm$^{-3}$ is the entropy density at present~\cite{Aghanim:2018eyx}.
Figure~\ref{fig:Ti-beta} shows with thick black lines the parameter space reproducing the observed DM density for different DM masses.
The shaded regions represent areas constrained by different observables: $\Min\lesssim 10^{-1}$~g and $\Min\gtrsim 2\times 10^8$~g are disfavored by CMB and BBN (both in red), $\beta$ values smaller than $\sim 10^{-24}$ can not accommodate the total observed DM abundance (green) (Eq.~\eqref{eq:betamin}), whereas large values for $\beta\gtrsim 10^{-2}$ lead to GW backreaction (Eq.~\eqref{eq:GW}), and $\beta\gtrsim 10^{-5}$ tend to produce hot DM (blue) (Eq.~\eqref{eq:WDMwithout}).
It is important to note that the latter constraint {\it only} applies to the case of light DM ($\mdm\ll \Tbh$), with mass $\mdm\lesssim 30$~GeV.
Finally, the dotted red line shows the transition between radiation (lower part) and matter-dominated eras (upper part).

In this figure, the effects of both the SM-DM interactions and the DM self-interactions have been neglected, and therefore it corresponds to DM produced solely via Hawking evaporation of PBHs.
The thick black lines show three different slopes, corresponding to three different regimes.
If PBHs dominated the universe energy density (above the red dotted line), the DM yield is independent of $\beta$, cf. Eq.~\eqref{eq:YdmMD}, and therefore the lines are vertical.
In this regime, $\mdm\simeq 10^9$~GeV is the lowest viable DM mass.
However, a $\beta$ dependence shows up if, during the whole BH lifetime, the universe was radiation dominated (below the dotted line), Eq.~\eqref{eq:YdmRD}.
In this case, two regimes arise, depending on whether DM is lighter or heavier than the initial BH temperature, Eq.~\eqref{eq:N}.
In the former case $\beta\propto\Tin$, whereas in the latter $\beta\propto\Tin^{-3}$.\\
In the present case where DM self-interactions are not efficient, DM has to be heavier than $\mathcal{O}(1)$~MeV in order not to be hot, and can be as heavy as $M_P$~\cite{Chung:1999ve, Giudice:1999fb} (notice that we are not considering a BH evaporation process stopping at $\Tbh\sim M_P$, with the associated production of Planck mass relics~\cite{MacGibbon:1987my, Barrow:1992hq, Carr:1994ar, Dolgov:2000ht, Baumann:2007yr,Hooper:2019gtx}).\\

Before closing the section, two comments are in order.
On the one hand, as mentioned previously, if one requires the PBHs to radiate the whole observed DM abundance, a lower limit on $\beta$ appears in the radiation-dominated era when $\mdm=\Tbhin$, and corresponds to
\begin{equation}\label{eq:betamin}
    \beta \geq \frac{4\pi^4}{45\zeta(3)\,\Cn}\frac{\gss(\Tin)}{g_\text{DM}}\frac{\mdm\,Y_\text{DM}}{\Tin}\simeq 7.6\times10^{-23} \left(\frac{\gs(\Tbh)}{106.75}\right)^{5/4} \left(\frac{0.2}{\gamma}\right)^{1/2} \left(\frac{\Min}{1~\text{g}}\right)^{1/2}.
\end{equation}
This bound is shown in green in Fig.~\ref{fig:Ti-beta}. 
On the other hand, we note that due to their large initial momentum, DM particles could have a large free-streaming length leading to a suppression on the structure formation at small scales.
In the present scenario where DM has no interactions with the SM or with itself, the DM momentum simply redshifts, and its value $p_0$ at present is~\cite{Fujita:2014hha}
\begin{equation}\label{eq:p0}
    p_0=\frac{a_\text{ev}}{a_0}p_\text{ev}
    =\frac{a_\text{ev}}{a_\text{eq}}\frac{a_\text{eq}}{a_0}p_\text{ev}
    =\frac{a_\text{ev}}{a_\text{eq}}\frac{\Omega_R}{\Omega_m}p_\text{ev}
    =\left[\frac{\gss(T_\text{eq})}{\gss(\Tev)}\right]^{1/3}\frac{T_\text{eq}}{\Tev}\frac{\Omega_R}{\Omega_m}p_\text{ev}\,,
\end{equation}
where $T_\text{eq}$ and $a_\text{eq}$ correspond to the temperature and the scale factor at the matter-radiation equality, respectively.
For light DM ($\mdm\ll\Tbhin$), $p_\text{ev}\simeq\Tbhin$ and by using Eq.~\eqref{eq:Tev}, the DM typical momentum at present can be estimated as
\begin{equation}
    p_0\simeq\left[\frac{\gss(T_\text{eq})}{\gss(\Tev)}\right]^{1/3}\left[\frac{10240}{9\,\gs(\Tbh)}\right]^{1/4}T_\text{eq}\frac{\Omega_R}{\Omega_m} \left[\frac{\Min}{M_P}\right]^{1/2}
    \simeq 7.8\times 10^{-14} \left[\frac{\Min}{M_P}\right]^{1/2}\text{GeV},
\end{equation}
where $T_\text{eq}\simeq 0.8$~eV, $\Omega_R\simeq 5.4\times 10^{-5}$ and $\Omega_m\simeq 0.315$~\cite{Aghanim:2018eyx, Tanabashi:2018oca} were used.
A lower bound on the DM mass can be obtained from the upper bound on a typical velocity of warm DM at present time.
Taking $v_\text{DM}\lesssim 1.8\times 10^{-8}$~\cite{Masina:2020xhk} for $\mdm\simeq 3.5$~keV~\cite{Irsic:2017ixq}, one gets 
\begin{equation}\label{eq:WDMwithout}
    \frac{\mdm}{1~\text{GeV}}\gtrsim 4\times 10^{-6}\left(\frac{\Min}{M_P}\right)^{1/2}\simeq 2\times 10^{-3}\left(\frac{\Min}{\text{g}}\right)^{1/2}.
\end{equation}
This bound is shown in blue in Fig.~\ref{fig:Ti-beta}, and {\it only} constrains light DM particles with mass $\mdm\lesssim 30$~GeV.

\section{Self-interactions}\label{sec:SI}
In the previous section, the production of {\it collisionless} DM particles via the evaporation of PBHs was presented.
However, DM can feature sizable {\it self-interactions}, dramatically changing its expected relic density.
To analytically understand the role played by DM self-interactions~\cite{Bernal:2020gzm}, let us study the DM production under the assumption of an instantaneous evaporation of the PBHs at $T=\Tev$.

\subsection{Radiation Dominated Universe}
First, we focus on the case where the universe was dominated by SM radiation during the whole lifetime of PBHs.

\subsubsection{Light DM}
In the case where the DM is lighter than the initial BH temperature ($\mdm\ll\Tbhin$), each BH radiates $\Ndm$ DM particles with a mean energy $\langle E\rangle= 6\,\Tbh$, Eq.~\eqref{eq:meanE}.
The total DM energy density radiated by a BH can be estimated by
\begin{equation}
    \rho_\text{DM}(T=\Tev) \simeq \ndm(T=\Tev)\,\Tbh
    = \beta\,\frac{\gdm\,\zeta(3)\,\Cn}{2\pi^2}\,\Tin\,\Tev^3\,,
\end{equation}
where the SM entropy conservation and Eq.~\eqref{eq:N} were used.
Let us notice that the produced DM population inherits the BH temperature, and, as DM is {\it not} in chemical equilibrium, develops a large chemical potential.

If the dark sector has sizable self-interactions guaranteeing that elastic scatterings within the dark sector reach kinetic equilibrium, DM thermalizes with a temperature $\Tp$ in general different from the SM one $T$ and the BH temperature $\Tbh$.
Thermalization within the dark sector is guaranteed if DM reaches kinetic equilibrium, i.e., if the rate of DM elastic scattering is higher than the Hubble expansion rate.
Assuming an instantaneous thermalization process,%
\footnote{Going beyond this approximation requires the solution of multiple momenta of the Boltzmann equation.}
and taking into account the instantaneous conservation of the DM energy density, the temperature $\Tpev$ in the dark sector just after thermalization is
\begin{equation}\label{eq:TpevlightRD}
    \Tpev =
    \begin{cases}
        \left(\beta\frac{15\zeta(3)\,\Cn}{\pi^4\,\Cr}\right)^{1/4}\Tin^{1/4}\,\Tev^{3/4}\qquad&\text{for}\quad \mdm\ll\Tpev\,,\\[8pt]
        \frac23\mdm\, \mathcal{W}^{-1}\left[\frac{1}{3\pi} \left(\frac{\gdm\,\mdm^4}{\rho_\text{DM}(T=\Tev)}\right)^{2/3} \right]\qquad&\text{for}\quad \mdm\gg\Tpev\,,
    \end{cases}
\end{equation}
for the cases $\mdm\ll\Tpev<\Tbhin$ and $\Tpev\ll\mdm<\Tbhin$, respectively, where $\mathcal{W}$ corresponds to the principal value of the Lambert function, and $\Cr = 1$ (bosonic DM) or $7/8$ (fermionic DM).
There is a net decrease of the DM mean kinetic energy $\Tpev/\Tbh\ll 1$, and therefore DM can become non-relativistic due to thermalization effects.

Additionally, if number-changing self-interactions within the dark sector reach chemical equilibrium, the DM number density just after thermalization is therefore
\begin{equation}\label{eq:ndmboostlightRD}
    \ndm(\Tpev) = 
    \begin{cases}
        \frac{15^{3/4}\zeta(3)^{7/4}\,\gdm\,\Cn^{7/4}}{\pi^5\,\Cr^{3/4}} \beta^{3/4} \left(\Tin\,\Tev^3\right)^{3/4}\qquad&\text{for}\quad \mdm\ll\Tpev\,,\\[10pt]
        \beta\,\frac{\gdm\,\zeta(3)\,\Cn}{2\pi^2}\,\frac{\Tin\,\Tev^3}{\mdm}\qquad&\text{for}\quad \mdm\gg\Tpev\,.
    \end{cases}
\end{equation}

The overall effect of self-interactions and in particular of number-changing interactions within the dark sector is to decrease the DM temperature, increasing the DM number density.
Such an increase can be characterized by a boost factor $B$ defined by comparing the DM number densities taking into account the case with relative to the case without thermalization in the dark sector:
\begin{equation}\label{eq:boostlightRD}
    B \equiv \frac{\ndm(\Tpev)}{\ndm(T=\Tev)}
    \simeq
    \begin{cases}
        \frac{2}{\beta^{1/4}\,\pi^3}\left(\frac{15\,\zeta(3)\,\Cn}{\Cr}\right)^{3/4} \frac{\Tbhin}{\Tin^{1/4}\,\Tev^{3/4}}\qquad&\text{for}\quad \mdm\ll\Tpev\,,\\[10pt]
        \frac{\Tbhin}{\mdm}\qquad&\text{for}\quad \mdm\gg\Tpev\,.
    \end{cases}
\end{equation}

\subsubsection{Heavy DM}
In the case where DM is heavier than the initial BH temperature ($\mdm\gg\Tbhin$), PBHs radiate DM particles with a mean energy $\langle E\rangle= 6\,\mdm$, Eq.~\eqref{eq:meanE}.
The total DM energy density radiated by a BH can be estimated by
\begin{equation}
    \rho_\text{DM}(T=\Tev) \simeq \beta\, \frac{\zeta(3)\,\gdm\,\Cn}{2\pi^2} \frac{\Min\,\mdm}{M_P^2} \Tin\,\Tev^3\,,
\end{equation}
and therefore, the temperature in the dark sector just after thermalization of non-relativistic DM particles is
\begin{equation}
    \Tpev \simeq \frac23\mdm\,{\mathcal W}^{-1}\left[\frac{1}{3\pi} \left(
    \frac{2\pi^2}{\beta\,\zeta(3)\,\Cn} \frac{\mdm^3\,M_P^2}{\Tin\,\Tev^3\,\Min} \right)^{2/3} \right],
\end{equation}
which this time corresponds to a mild decrease of the mean DM kinetic energy $\Tpev/\mdm \lesssim 1$.
The DM number density just after thermalization for non-relativistic DM particles is
\begin{equation}
    \ndm(\Tpev) = \beta\,\frac{\gdm\,\zeta(3)\,\Cn}{2\pi^2}\,\frac{\Min}{M_P^2}\,\Tin\,\Tev^3\,,
\end{equation}
and hence a boost factor
\begin{eqnarray}\label{eq:boostheavytRD}
    B \simeq 1\,.
\end{eqnarray}
We notice that this factor coincides with the one for DM heavier than $\Tpev$ in Eq.~\eqref{eq:boostlightRD}, by taking the limit $\mdm\to\Tbhin$.
A boost factor $B\simeq 1$ (i.e., no boost!) was expected in this case where the originally Hawking radiated particles were almost non-relativistic.

\subsection{Matter Dominated Universe}
After having studied the case where the universe was dominated by SM radiation during the whole lifetime of the PBH, in this section we focus on the other scenario, in which PBHs dominated the energy density.
In this section, an analysis analogous to the one presented previously will be followed.

\subsubsection{Light DM}
The total DM energy density for light particles ($\mdm\ll\Tbhin$) radiated by a BH is
\begin{equation}
    \rho_\text{DM}(T=\Tbev) \simeq \frac{\gdm\,\zeta(3)\,\Cn}{2\pi^2}\,\Tbev^4\,.
\end{equation}
Again, assuming an instantaneous thermalization process, and taking into account the instantaneous conservation of the DM energy density, the temperature $\Tpev$ in the dark sector just after thermalization
\begin{equation}\label{eq:pDMlightMD}
    \Tpev \simeq
    \begin{cases}
        \left(\frac{15\,\zeta(3)\,\Cn}{\pi^4\,\Cr}\right)^{1/4} \Tbev\qquad&\text{for}\quad \mdm\ll\Tpev\,,\\[8pt]
        \frac23\mdm\, \mathcal{W}^{-1}\left[\frac{1}{3\pi} \left(\frac{\gdm\,\mdm^4}{\rho_\text{DM}(T=\Tbev)}\right)^{2/3} \right]\qquad&\text{for}\quad \mdm\gg\Tpev\,,
    \end{cases}
\end{equation}
the DM number density just after thermalization
\begin{equation}
    \ndm(\Tpev) \simeq
    \begin{cases}
        \frac{\zeta(3)\,\Cn}{\pi^2}\gdm\left(\frac{15\,\zeta(3)\,\Cn}{\pi^4\,\Cr}\right)^{3/4} \Tbev^3\qquad&\text{for}\quad \mdm\ll\Tpev\,,\\[8pt]
        \frac{\gdm\,\zeta(3)\,\Cn}{2\pi^2}\,\frac{\Tbev^4}{\mdm}\qquad&\text{for}\quad \mdm\gg\Tpev\,,
    \end{cases}
\end{equation}
and therefore the boost factor becomes
\begin{equation}\label{eq:boostlightMD}
    B \simeq
    \begin{cases}
        \frac{2}{\pi^3} \left(\frac{15\,\zeta(3)\,\Cn}{\Cr}\right)^{3/4} \frac{\Tbhin}{\Tbev}\qquad&\text{for}\quad \mdm\ll\Tpev\,,\\[8pt]
        \frac{\Tbhin}{\mdm}\qquad&\text{for}\quad \mdm\gg\Tpev\,,
        \end{cases}
\end{equation}
which matches the result for the radiation domination case, Eq.~\eqref{eq:boostlightRD}, in the limit $\beta\to\Tev/\Tin$ with $\Tbev=\Tev$.

\subsubsection{Heavy DM}
The total DM energy density for heavy particles ($\mdm\gg\Tbhin$) radiated by a BH is
\begin{equation}
    \rho_\text{DM}(T=\Tbev) \simeq \frac{\gdm\,\zeta(3)\,\Cn}{2\pi^2}\,\Tbev^4\frac{\mdm}{\Tbhin}\,.
\end{equation}
The temperature $\Tpev$ in the dark sector just after thermalization of non-relativistic DM particles
\begin{equation}
    \Tpev \simeq
    \frac23\mdm\, \mathcal{W}^{-1}\left[\frac{1}{3\pi} \left(\frac{\gdm\,\mdm^4}{\rho_\text{DM}(T=\Tbev)}\right)^{2/3} \right],
\end{equation}
the DM number density just after thermalization for non-relativistic DM particles becomes
\begin{equation}
    \ndm(T=\Tbev) \simeq \frac{\gdm\,\zeta(3)\,\Cn}{2\pi^2}\,\frac{\Tbev^4}{\Tbhin}\,,
\end{equation}
and hence the boost factor
\begin{equation}\label{eq:boostheavyMD}
    B \simeq 1\,,
\end{equation}
which matches the results for the radiation dominated case, Eq.~\eqref{eq:boostheavytRD}, and for light DM in matter domination, Eq.~\eqref{eq:boostlightMD}, in the limit $\mdm\to\Tbhin$.\\

\begin{figure}
	\centering
	\includegraphics[scale=0.58]{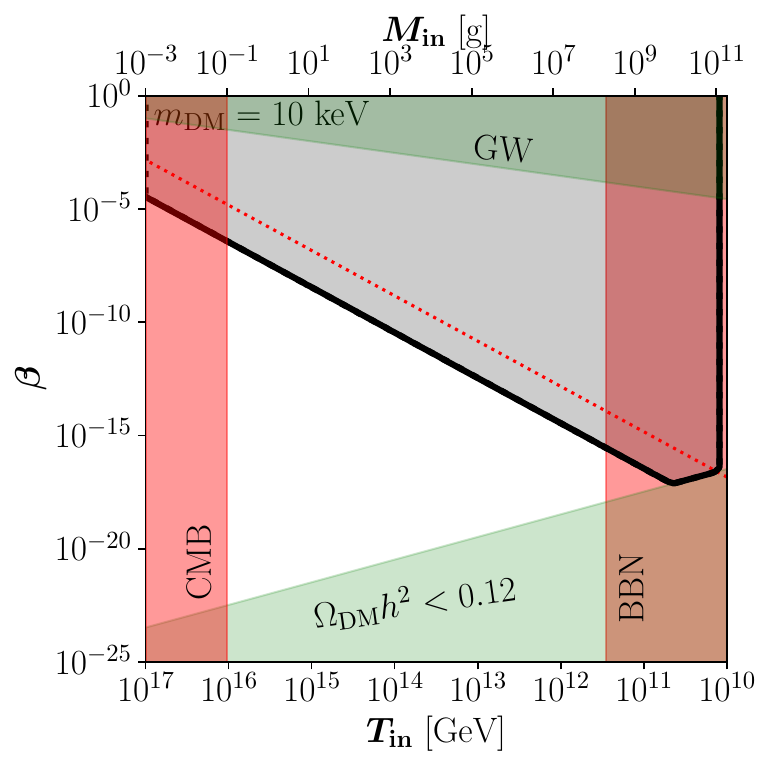}
	\includegraphics[scale=0.58]{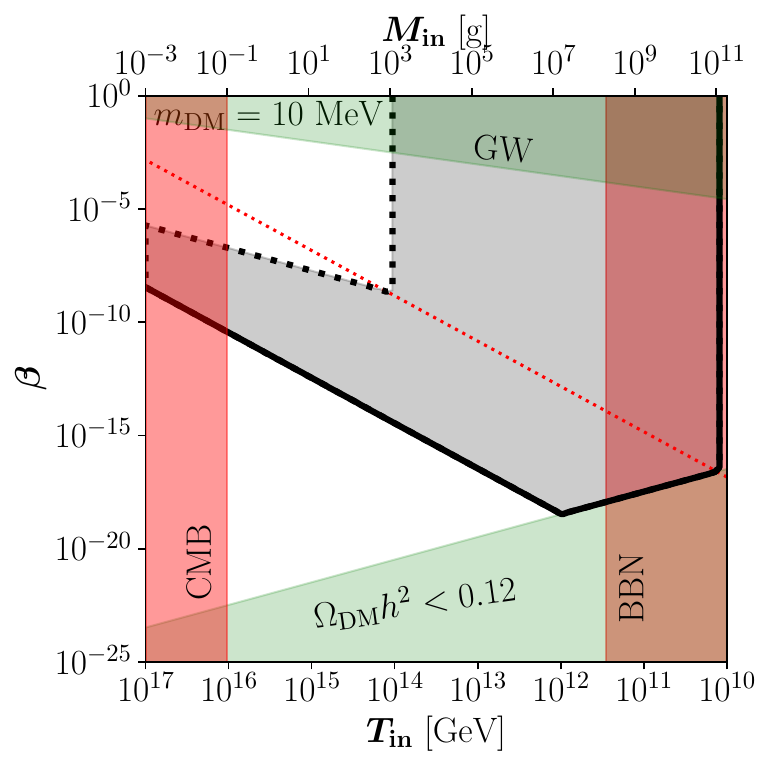}
	\includegraphics[scale=0.58]{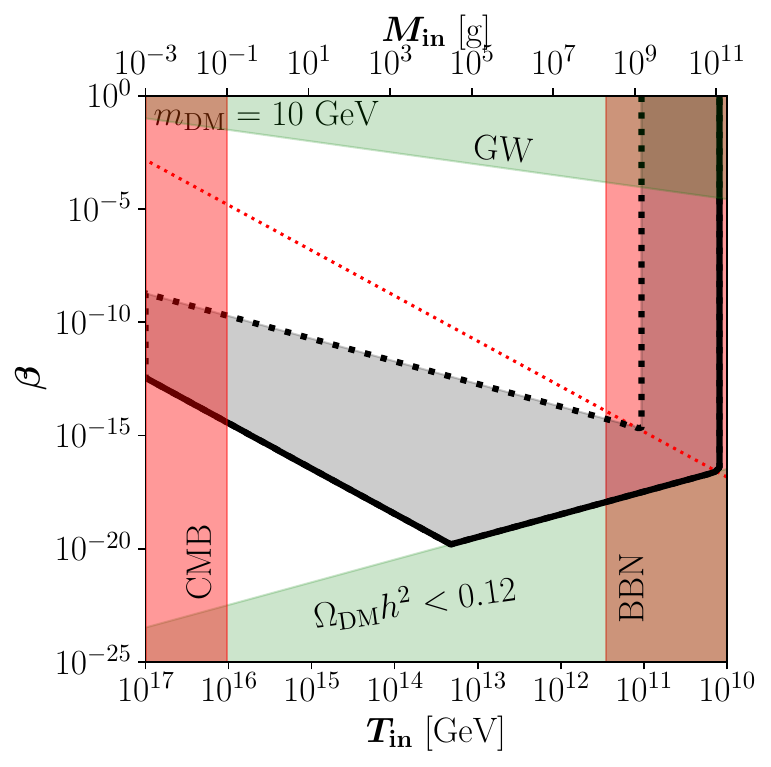}
	\includegraphics[scale=0.58]{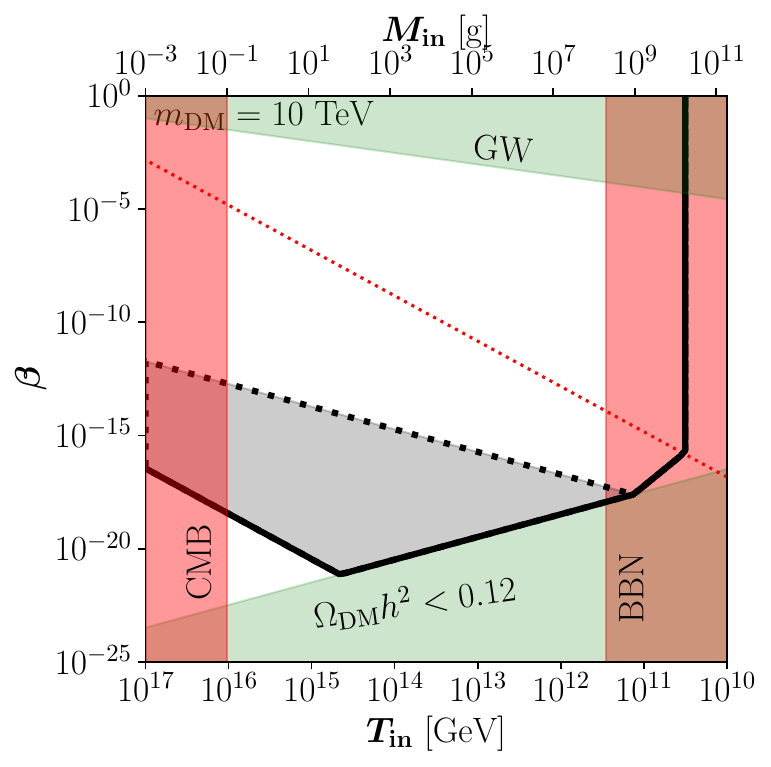}
	\caption{Parameter space reproducing the observed DM abundance (shaded gray areas) from PBH evaporation, for different DM masses.
	The thick black lines show the limiting cases without (dotted lines) and with a maximum effect from DM self-interactions (solid lines).
	The shaded green and red areas are excluded by different observables and described in the text.}
	\label{fig:Ti-beta-masses}
\end{figure}

The impact of DM self-interactions is shown in Fig.~\ref{fig:Ti-beta-masses}, for different DM masses.
Dotted and solid thick lines correspond to the limiting cases without and with a maximal effect from self-interactions, respectively, in the same parameter space used in Fig.~\ref{fig:Ti-beta}.
Out of the six regimes presented previously, four are visible in the plots and are described in ascending order for $\Tin$.
\begin{itemize}
    \item The observed DM abundance can be generated in the case where the PBHs dominated the universe energy density (above the red dotted line), only for heavy DM, i.e. $\mdm > \Tbhin$.
    The DM yield is independent from $\beta$ (Eq.~\eqref{eq:YdmMD}) and there is no boost due to self-interactions (Eq.~\eqref{eq:boostheavyMD}), as can be seen in the lower right panel corresponding to $\mdm=10$~TeV.
    We notice that this scenario is typically excluded by the BBN constraint, and only viable for $\mdm\gtrsim 10^9$~GeV. 
    \item The case of heavy DM, this time in a radiation dominated scenario, appears when $\mdm\gtrsim 100$~GeV.
    This case is visible in the lower right panel, where the DM yield is given by Eq.~\eqref{eq:YdmRD} without a significant boost, Eq.~\eqref{eq:boostheavytRD}. It follows that $\beta\propto\Tin^{-3}$.
    This scenario is again typically in tension with the BBN observations, and only viable when $\mdm\gtrsim 10^5$~GeV.
    \item The third regime corresponds to $\Tpev\ll\mdm\ll\Tbhin$, where the DM yield in Eq.~\eqref{eq:YdmRD} is boosted by a factor $\Tbhin/\mdm$, Eq.~\eqref{eq:boostlightRD}.
    In this case, $\beta \propto \Tin^{-1}$.
    We notice that values of $\beta$ smaller than the ones required in this scenario always produce a DM underabundance.
    \item The last regime happens for light DM, $\mdm\ll\Tpev$ in a radiation-dominated universe.
    The DM yield in Eq.~\eqref{eq:YdmRD} is boosted by a factor $\propto \Tbhin/(\Tin^{1/4}\,\Tev^{3/4})$, Eq.~\eqref{eq:boostlightRD}.
    In this case, the DM abundance requires $\beta \propto \Tin^2$.
\end{itemize}

We would like to emphasize that the current boost factors have to be understood as the {\it maximum} increase of the DM number density due to self-interactions.
They can be reached if the number-changing DM self-interactions freeze-out while DM is relativistic (or soon after chemical equilibrium is achieved), to avoid a DM depletion due to cannibalization processes, e.g. $3\to2$ or $4\to2$ annihilations.
Furthermore, we notice that in specific models, chemical equilibrium in the dark sector may not be achieved due to the perturbativity limit, and therefore the maximum boost can not be attained.
However, going further from these general considerations requires the choice of a specific particle physics model, see, e.g., Refs.~\cite{Bernal:2020gzm, March-Russell:2020nun, Bernal:2015ova, Bernal:2015xba, Bernal:2017mqb, Chu:2013jja, Falkowski:2017uya, Herms:2018ajr, Heeba:2018wtf, Mondino:2020lsc} for concrete realizations of this scenario, although not in the context of PBHs.\\

Additionally to the increase of the DM number density, self-interactions also reduce the DM typical momentum.
The bound on the DM mass coming from the possible suppression on the structure formation rate due to its free-streaming length is therefore eased.
In the case with self-interactions, the momentum of DM particles in a matter-dominated scenario is $p_\text{ev}\simeq \Tpev$, see Eq.~\eqref{eq:pDMlightMD}.
Therefore, from Eq.~\eqref{eq:p0} one has that
\begin{equation}
    p_0 \simeq \left[\frac{\gss(T_\text{eq})}{\gss(\Tev)}\right]^{1/3} T_\text{eq}\, \frac{\Omega_R}{\Omega_m} \left(\frac{15\,\zeta(3)\,\Cn}{\pi^4\,\Cr}\right)^{1/4},
\end{equation}
for light DM with $\mdm\ll\Tpev$, and assuming that DM momentum scales like $a^{-1}$.
This can be achieved if $p$ simply redshifts, i.e., if DM kinetically decouples just after thermalization, or if the kinetic equilibrium is broken when DM is still relativistic, and implies a lower limit on the DM mass
\begin{equation}\label{eq:WDMwith}
    \mdm \gtrsim 4~\text{keV}\,,
\end{equation}
following the same procedure as for the case without self-interactions, in section~\ref{sec:DM}.
However, if chemical equilibrium is active when DM is non-relativistic, number-changing interactions enforce DM temperature to fall only logarithmically until they freeze-out~\cite{Carlson:1992fn}.
These cannibalization processes rise the DM temperature relative to the SM, increasing the DM momentum and therefore strengthening the bound on the DM mass.
In that sense, Eq.~\eqref{eq:WDMwith} has to be understood as a minimal lower bound.

\section{Conclusions} \label{sec:con}

DM production via Hawking evaporation of PBHs constitutes an irreducible process in the early universe.
This channel can be dominant, for instance, if the dark and visible sectors are disconnected.
In that case, after production the DM comoving density stays constant until today.
Additionally, light DM is radiated relativistically and could erase small-scale structures via free-streaming.
This enforces DM to be heavier than a few MeV.

However, this paradigm is modified if DM features sizable self-interactions.
Thermalization and number-changing processes in the dark sector can have strong impacts, in particular enhancing the DM relic abundance by several orders of magnitude.
In this paper we have estimated the boost from general arguments such as the conservation of energy and entropy, independently from the underlying particle physics details of the dark sector.
Two main consequences can be highlighted:
$i)$ As the DM abundance is increased, a smaller initial energy density of PBHs (encoded in the parameter $\beta$) is required.
$ii)$ Thermalization in the dark sector decreases the mean DM kinetic energy, relaxing the bound from structure formation and hence, allowing for lighter DM in the keV ballpark.

We note that this work focused on a {\it model-independent} analysis, that applies to very general DM models.
For instance, the results presented in Fig.~\ref{fig:Ti-beta-masses} should be understood as the maximum boost that DM self-interactions can produce. In particular, any DM model coming from a simple hidden sector (more elaborate scenarios featuring multicomponent DM, bound states, mediators, etc, clearly escape from this approach) should lie between the dotted line (no DM self-interactions) and the solid line (maximum effect from DM self-interactions).
Finally, we emphasize that the analysis of this work could naturally be reproduced for specific particle physics models, allowing to take into account all particle physics, cosmological and astrophysical constraints, and to do a more elaborated treatment for the the DM thermalization.

Before concluding, we note that thermalization and number-changing interactions naturally appear in scenarios where DM features sizable self-interactions.
Those DM self-interactions could play a role in the solution of the so-called `core vs. cusp'~\cite{Flores:1994gz, Moore:1994yx, Oh:2010mc, Walker:2011zu} and `too-big-to-fail' problems~\cite{BoylanKolchin:2011de, BoylanKolchin:2011dk, Garrison-Kimmel:2014vqa, Papastergis:2014aba} arising at small scales.
For this to be the case, the required self-scattering cross section over DM mass needs to be of the order of 0.1--2~cm$^2$/g at the scale of dwarf galaxies~\cite{Kaplinghat:2015aga, Fry:2015rta}, and smaller than $1.25$~cm$^2$/g~\cite{Randall:2007ph} or even 0.065~cm$^2$/g~\cite{Andrade:2020lqq} at the scale of galaxy clusters.
Additionally, possible imprints may arise from the evaporation of PBHs before the BBN onset. Since gravitons are always within the BH evaporation products and do not reach thermal equilibrium, there would be a corresponding stochastic background of gravitational waves with a frequency in the range  above $10$~THz~\cite{BisnovatyiKogan:2004bk, Anantua:2008am, Dolgov:2011cq}, which is high enough to be within the sensitivity region of ongoing and near future gravity wave experiments~\cite{Morrison:2018xla, Ito:2019wcb, Ejlli:2019bqj, Zagorac:2019ekv, Inomata:2020lmk, Hooper:2020evu, Domcke:2020yzq}.

\section*{Acknowledgments}

The authors thank Xiaoyong Chu and members of ``El Journal Club más sabroso'' for fruitful discussions.
NB received funding from Universidad Antonio Nariño grants 2018204, 2019101, and 2019248, the Spanish MINECO under grant FPA2017-84543-P, and the Patrimonio Autónomo - Fondo Nacional de Financiamiento para la Ciencia, la Tecnología y la Innovación Francisco José de Caldas (MinCiencias - Colombia) grant 80740-465-2020.
The work of OZ is supported by Sostenibilidad-UdeA, the UdeA/CODI Grant 2017-16286, and by COLCIENCIAS through the Grant 111577657253.  
This project has received funding /support from the European Union's Horizon 2020 research and innovation programme under the Marie Skłodowska-Curie grant agreement No 860881-HIDDeN.

\bibliographystyle{JHEP}
\bibliography{biblio}

\end{document}